\documentclass[aps,prd,reprint,onecolumn, groupedaddress,longbibliography, 12pt]{revtex4-1}
\usepackage{amsmath,amssymb}
\usepackage{graphicx}
\usepackage{color}
\usepackage{caption}
\usepackage{hyperref}
\usepackage{graphicx}
\begin{document}

\title{Analogue black holes in relativistic BECs: \\ Mimicking Killing and universal horizons\\[10pt]}

\author{Bethan Cropp}
\email[]{bcropp@issertvm.ac.in}
\affiliation{SISSA, Via Bonomea 265, 34136 Trieste, Italy, \\
     INFN sezione di Trieste, Via Valerio 2, 34127 Trieste, Italy
                       and \\
School of Physics, Indian Institute of Science Education and Research Thiruvananthapuram (IISER-TVM), Trivandrum 695016, India.}
\author{Stefano Liberati}
\email[]{stefano.liberati@sissa.it}
\affiliation{SISSA, Via Bonomea 265, 34136 Trieste, Italy
                                       and \\
     INFN sezione di Trieste, Via Valerio 2, 34127 Trieste, Italy.}
\author{Rodrigo Turcati}
\email[]{rturcati@sissa.it}
\affiliation{SISSA, Via Bonomea 265, 34136 Trieste, Italy	
                                       and \\
     INFN sezione di Trieste, Via Valerio 2, 34127 Trieste, Italy.}

\def\d{{\mathrm{d}}}
\newcommand{\scri}{\mathscr{I}}
\newcommand{\sun}{\ensuremath{\odot}}
\def\J{{\mathscr{J}}}
\def\L{{\mathscr{L}}}
\def\sech{{\mathrm{sech}}}
\def\T{{\mathcal{T}}}
\def\tr{{\mathrm{tr}}}
\def\diag{{\mathrm{diag}}}
\def\ln{{\mathrm{ln}}}
\def\Horava{Ho\v{r}ava}
\def\Aether{\AE{}ther}
\def\AEther{\AE{}ther}
\def\aether{\ae{}ther}
\def\UH{{\text{\sc uh}}} 
\def\KH{{\text{\sc kh}}}

\begin{abstract}

\noindent Relativistic Bose--Einstein condensates (rBECs) have recently become a well-established system for analogue gravity. Indeed, while such relativistic systems cannot be yet realized experimentally, they provide an interesting framework for mimicking metrics for which no analogue is yet available, so paving the way for further theoretical and numerical explorations. In this vein, we here discuss black holes in rBECs and explore how their features relate to the bulk properties of the system. We then propose the coupling of external fields to the rBEC as a way to mimic non-metric features. In particular, we use a Proca field to simulate an \aether\ field, as found in Einstein--\Aether\ or \Horava--Lifshitz gravity. This allows us to mimic a universal horizon, the causal barrier relevant for superluminal modes in these modified gravitational theories.


\end{abstract}

\maketitle

\tableofcontents

\section{Introduction}

Over the past few decades, there has been a growing interest in using condensed matter systems as toy models to investigate kinematical features of classical and quantum field theories in curved spacetimes \cite{Barcelo:2005fc}. The main idea behind the {\it analogue gravity} formalism is the following: when the equations of fluid dynamics are linearized under appropriate conditions, the phonon propagation can be described by a relativistic equation of motion in a curved spacetime, the emergent geometry being described by the acoustic metric. Among many of these models, one of the most explored condensed matter systems are the Bose--Einstein condensates (BECs) \cite{Garay:1999sk,Garay:2000jj}. 

In BECs, one can split the classical ground state from its quantum fluctuations, where the massless excitations on the condensate can be described in the same way as a scalar field propagating in a curved spacetime. Therefore, BECs provide an interesting framework to probe issues of quantum field theories in curved backgrounds. For the most part, these investigations have considered nonrelativistic BECs. Nevertheless, an acoustic description of the phonons on the top of the condensate taking into account a relativistic BEC (rBEC) has been found recently \cite{Fagnocchi:2010sn}. While this system is not yet realized in laboratory experiments, it provides a richer framework for mimicking more general spacetimes than those describable by nonrelativistic BEC (or fluids). As such, it is an ideal testbed for numerical and theoretical studies on the interplay between phenomena such as Hawking radiation and the structure of an emergent spacetime.

The purpose of this paper is a twofold investigation of analogue black holes in rBEC systems. To date, the only explicit analogue spacetime developed with rBECs is the $k=-1$ Friedmann-Lema\^itre-Robertson-Walker cosmological solution \cite{Fagnocchi:2010sn}. As black holes are some of the most useful and simple analogue spacetimes, we investigate black holes and how to relate their features to the properties of the rBEC. We look explicitly at modeling the Schwarzschild solution, and examine a simplified flow leading to a canonical acoustic black hole. 

The second aim is to build a model in which the universal horizon emerges in the context of analogue gravity. Universal horizons appear in some theories with Lorentz violation such as Einstein--{\AE}ther and \Horava--Lifshitz gravity. In these theories, particles can propagate superluminally, and potentially with infinite velocity, making the Killing horizon causally irrelevant, at least at high energies. However, the concept of causality remains, as particles must move forward through surfaces of constant khronon time. Therefore the notion of horizon holds in this framework, the universal horizon being the surface where the khronon time foliation is a surface of constant radius, so to move forward in time a particle must move inwards. Thus the universal horizon is the true causal barrier in such spacetimes. 

Introducing the \aether\ field in the context of the analogue gravity may shed some light on issues related to the emission of Hawking radiation in these Lorentz-violating gravity theories. Recently, it was suggested in Refs. \cite{Berglund:2012fk,Cropp:2013sea} that the presence of the universal horizon modifies the boundary conditions for the modes, so the Hawking radiation would be generated not at the Killing horizon, but in the universal horizon. Nevertheless, in Ref. \cite{Michel:2015rsa} it was claimed that the thermal emission arises from the Killing horizon with a temperature fixed by its surface gravity. 

Now, owing to the great interest generated by these special spacelike hypersurfaces, it would be useful to understand them from the point of view of analogue gravity. The metric felt by perturbations, as we will explicitly demonstrate in the next section, is identical to the relativistic acoustic geometry previously studied in Refs. \cite{Bilic:1999sq, Visser:2010xv}. One may wonder if it is possible to mimic a universal horizon using such a geometry and the preferred frame defined by the four-velocity. It was explicitly shown in Ref. \cite{Cropp:2013sea}, that using the standard nonrelativistic flow, one cannot mimic a universal horizon. One might similarly try with the relativistic acoustic metric~\cite{Visser:2010xv},
\begin{equation}
G_{\mu\nu}=\rho\frac{c}{c_{s}}\left[g_{\mu\nu}+\left(1-\frac{c_s^2}{c^2}\right)\frac{v_{\mu}v_{\nu}}{c^2}\right],
\end{equation}
where $G_{\mu\nu}$ is the effective metric, $g_{\mu\nu}$ is the background metric (which we pick to be Minkowski, as appropriate for any laboratory setup) and $g_{\mu\nu}v^{\mu}v^{\nu}=-c^2$. If we want to model an \AE\ black hole, we have a timelike Killing vector $\chi$ and using the flow four-velocity as the candidate \aether\ field, the condition for a universal horizon is
\begin{equation}
\chi^{\mu}v^{\nu}G_{\mu\nu}=\rho\frac{c}{c_{s}}\left[-v^{0}+\left(1-\frac{c_s^2}{c^2}\right)\frac{v^{2}v_{0}}{c^2}\right]=0,
\end{equation}
where using $v^{2}=-c^{2}$ in the above relation gives us the relation
\begin{equation}
v^{0}=\left(1-\frac{c_s^2}{c^2}\right)v^{0},
\end{equation}
which, as in the nonrelativistic case, is impossible to satisfy. 

It is clear that we truly need to add some new degrees of freedom to model the {\ae}ther. A possible way to achieve this is coupling the scalar field minimally to electromagnetism. In this way, one can interpret the fluid flow as the components of the acoustic metric and the {\aether} field can be associated to the gauge field. We will explore this scenario. However, due to the possibility of spurious gauge effects, we will see that the mentioned approach is inadequate. This leads us to consider a more appropriate choice in the form of a coupling to a massive vector field. This choice removes issues related to the gauge invariance allowing a clear interpretation of the \aether\ in the context of analogue gravity and giving us the freedom to mimic the universal horizon.

The paper is arranged as follows: We start by introducing and briefly rederiving the analogue metric of the rBEC. We then, in Sec. \ref{analogueBHs}, discuss fitting the Schwarzschild solution to the acoustic metric, and discuss a simplified black hole solution. We examine the condition for existence of a Killing horizon and determine the surface gravity in the variables of the condensate. After a brief introduction to universal horizons in Sec. \ref{universal}, we introduce the external field in Sec. \ref{analogueuh}, and then discuss the difficulties of modeling the known exact black hole solutions, and demonstrate that we can model the universal horizon in a simplified black hole, using the appropriate Proca field configuration. 

In our conventions the signature of the metric is $(-,+,+,+)$.


\section{Relativistic BECs: a new system for analogue black holes}\label{uncoupledrbec}

In this section we will give a brief review of the relativistic Bose--Einstein condensation and how it leads, in the appropriate hydrodynamical limit, to quasiparticles propagating over an emergent acoustic metric. For further details, see Ref. \cite{Fagnocchi:2010sn}. We start with the Lagrangian density for a complex scalar field $\phi(\mathbf{x},t)$, given by 
\begin{eqnarray}\label{lrbec}
\mathcal{L}=-\eta^{\mu\nu}\partial_{\mu}\phi^{*}\partial_{\nu}\phi-\left(\frac{m^{2}c^{2}}{\hbar^{2}}+V(t,\mathbf{x})\right)\phi^{*}\phi-U(\phi^{*}\phi;\lambda_{i}),
\end{eqnarray}
where $m$ is the mass of bosons, $V(t,\mathbf{x})$ is an external potential, $c$ is the speed of light, $U$ is a self-interaction term and $\lambda_{i}(t,\mathbf{x})$ are the coupling constants. 

The Lagrangian (\ref{lrbec}) is invariant under the global $U(1)$ symmetry and has a conserved current 
\begin{equation}
j^{\mu}=i(\phi^{*}\partial^{\mu}\phi-\phi\partial^{\mu}\phi^{*}), 
\end{equation}
which is related to a conserved ensemble charge $N-\bar{N}$, where $N(\bar{N})$ is the number of bosons (antibosons).

In the case where there are no self-interactions ($U=0$) and no external potential ($V=0$), the average number of bosons $n_{k}$ in the state of energy $E_{k}$ can be written as
\begin{equation}\label{nb}
N-\bar{N}=\Sigma_{k}[n_{k}-\bar{n}_{k}], 
\end{equation}
where 
\begin{equation}
n_{k}(\mu,\beta)=1/\left\lbrace{\exp}[\beta(|E_{k}|-\mu)]-1\right\rbrace, \qquad  \bar{n}_{k}(\mu,\beta)=1/\left\lbrace{\exp[\beta(|E_{k}|+\mu)]-1}\right\rbrace,
\end{equation}
and $\mu$ is the chemical potential, $T\equiv1/(k_{B}\beta)$ is the temperature and the energy of the state $k$ is given by $E^{2}_{k}=\hbar^{2}k^{2}c^{2}+m^{2}c^{4}$.

In a system of volume $\mathcal{V}$, the relation between the conserved charge density, $n=(N-\bar{N})/\mathcal{V}$, and the critical temperature is
\begin{equation}\label{nd}
n=C\int^{\infty}_{0}\d kk^{2}\frac{\sinh(\beta_{c}mc^{2})}{\cosh(\beta_{c}|E_{k}|)-\cosh(\beta_{c}mc^{2})}, 
\end{equation}
where $C=1/(4\pi^{3/2}\Gamma(3/2))$.

The nonrelativistic and ultrarelativistic limits can be obtained directly from equation (\ref{nd}). When $k_{B}T_{c}\ll{mc^{2}}$, the nonrelativistic limit is given by
\begin{equation}
k_{B}T_{c}=\frac{2\pi\hbar^{2}}{n}\left(\frac{n}{\zeta(3/2)}\right)^{2/3}, 
\end{equation}
where $\zeta$ is the Riemann zeta function. The ultra-relativistic limit is characterized by the condition $k_{B}T_{c}\gg{mc^{2}}$, which implies that
\begin{equation}
(k_{B}T_{c})^{2}=\frac{\hbar^{3}c\Gamma(3/2)(2\pi)^{3}}{4m\pi^{3/2}\Gamma(3)\zeta(2)}n.
\end{equation}

The condensation of the relativistic Bose gas occurs when $T\ll{T_{c}}$, where, using the mean-field approximation, the dynamics of the condensate is described by the nonlinear Klein--Gordon equation
\begin{equation}\label{groundstateequation}
\Box\phi-\left(\frac{m^{2}c^{2}}{\hbar^{2}}+V\right)\phi-U'\phi=0. 
\end{equation}

In this phase, it is possible to uncouple the BEC ground state from its perturbations. To perform this split one can insert $\phi=\varphi(1+\psi)$ in Eq. (\ref{groundstateequation}), where $\varphi$ is the classical background field satisfying the equation
\begin{equation}\label{nlkg}
\Box\varphi-\left(\frac{m^{2}c^{2}}{\hbar^{2}}+V\right)\varphi-U'\varphi=0, 
\end{equation}
and $\psi$ is a quantum relative fluctuation (i.e., of order $\hbar$). The modified Klein-Gordon equation (\ref{nlkg}) gives the dynamics of the ground state of the relativistic condensate.

It is also convenient decompose the degrees of freedom of the complex scalar classical field in terms of the Madelung representation, $\varphi=\sqrt{\rho}e^{i\theta}$. Using this prescription, the continuity equation and the condensate equation (\ref{nlkg}) assume the form
\begin{eqnarray}\label{cemrnc}
\partial_{\mu}(\rho{u^{\mu}})&=&0, \\
-u_{\mu}u^{\mu}&=&c^{2}+\frac{\hbar^{2}}{m^{2}}\left[V(x^{\mu})+U'(\rho;\lambda_{i}(x^{\mu}))-\frac{\Box\sqrt{\rho}}{\rho}\right],
\end{eqnarray}
where 
\begin{equation}\label{hypersurface}
u^{\mu}=\frac{\hbar}{m}\partial^{\mu}\theta
\end{equation}
is interpreted as the un-normalized four-velocity of the condensate.

The quantum perturbation $\psi$ satisfies 
\begin{eqnarray}\label{lpe}
\left\lbrace\left[i\hbar{{u}^{\mu}}\partial_{\mu}+T_{\rho}\right]\frac{1}{c_{0}^{2}}\left[-i\hbar{{u}^{\nu}}\partial_{\nu}+T_{\rho}\right]-\frac{\hbar^{2}}{\rho}\eta^{\mu\nu}\partial_{\mu}\rho\partial_{\nu}\right\rbrace\psi=0,
\end{eqnarray}
where $c_{0}^{2}\equiv\frac{\hbar^{2}}{2m^{2}}\rho{U''}$ is related to the interaction strength and 
\begin{equation}
T_{\rho}\equiv-\frac{\hbar^{2}}{2m}(\Box+\eta^{\mu\nu}\partial_{\mu}ln\rho\partial_{\nu})
\end{equation}
is a generalized kinetic operator. Although $c_{0}$ has dimension of velocity, it is not the speed of sound of the relativistic condensate. Nevertheless, as we will see in the relation (\ref{speedofsound}), both variables are connected.

Equation (\ref{lpe}) is the relativistic generalization of the Bogoliubov--de Gennes equation. The dispersion relation associated to Eq.~\eqref{lpe} has several limiting cases of interest which were fully explored in \cite{Fagnocchi:2010sn}. Here, we are particularly interested in the low-momentum regime which is characterized by the condition
\begin{equation}\label{analoguecondition}
|k|\ll\frac{mu^{0}}{\hbar}\left[1+\left(\frac{c_{0}}{u^{0}}\right)^{2}\right].
\end{equation}

Considering the phononic regime of the low-momentum range defined by Eq.~(\ref{analoguecondition}) and assuming that the background quantities $u$, $\rho$ and $c_{0}$ vary slowly (eikonal approximation) in space and time on scales comparable with the wavelength of the perturbation, i.e.,
\begin{eqnarray}
\left|\frac{\partial_{t}\rho}{\rho}\right|\ll{w}, \quad\quad \left|\frac{\partial_{t}c_{0}}{c_{0}}\right|\ll{w}, \quad\quad \left|\frac{\partial_{t}u_{\mu}}{u_{\mu}}\right|\ll{w}, 
\end{eqnarray}
one can disregard the quantum potential $T_{\rho}$, implying that the quasiparticles can be described in terms of an effective acoustic metric. Let us then show that the acoustic description can be achieved at the aforementioned scales. Applying the above assumptions, it is easy to see that Eq.~(\ref{lpe}) reduces to
\begin{eqnarray}\label{lpnqp}
\left[{{u}^{\mu}}\partial_{\mu}\left(\frac{1}{c_{0}^{2}}{{u}^{\nu}}\partial_{\nu}\right)-\frac{1}{\rho}\eta^{\mu\nu}\partial_{\mu}\left(\rho\partial_{\nu}\right)\right]\psi=0.
\end{eqnarray}

Now, in order to arrive at the acoustic metric, one can make use of the continuity equation (\ref{cemrnc}) and rewrite Eq.~(\ref{lpnqp}) as
\begin{eqnarray}\label{phononicequation}
\partial_{\mu}\left[-\rho\eta^{\mu\nu}+\frac{\rho}{c_{0}^{2}}u^{\mu}u^{\nu}\right]\partial_{\nu}\psi=0.
\end{eqnarray}

Equation (\ref{phononicequation}) can be expressed as
\begin{equation}
\partial_{\mu}\left(\gamma^{\mu\nu}\partial_{\nu}{\psi}\right)=0, 
\end{equation}
where $\gamma^{\mu\nu}$ is 
\begin{equation}
\gamma^{\mu\nu}=\frac{\rho}{c_{0}^{2}}\left(c_{0}^{2}\eta^{\mu\nu}-u^{\mu}u^{\nu}\right).
\end{equation}

Identifying $\gamma^{\mu\nu}=\sqrt{-g}g^{\mu\nu}$, 
\begin{equation}
\sqrt{-g}=\rho^{2}\sqrt{1-u^{\alpha}u_{\alpha}/c_{0}^{2}}, 
\end{equation}
and
\begin{eqnarray}
g^{\mu\nu}=\frac{1}{\rho{c}_{0}^{2}\sqrt{1-u^{\alpha}u_{\alpha}/c_{0}^{2}}}\left(c_{0}^{2}\eta^{\mu\nu}-u^{\mu}u^{\nu}\right).
\end{eqnarray}

Therefore, one can cast Eq.~(\ref{phononicequation}) in the form  
\begin{equation}\label{dalembertian}
\triangle{\psi}\equiv\frac{1}{\sqrt{-g}}\partial_{\mu}\left(\sqrt{-g}g^{\mu\nu}\partial_{\nu}{\psi}\right),
\end{equation}
which is a d'Alembertian for a massless scalar in a curved background. Inverting $g^{\mu\nu}$, one can then see that the acoustic metric $g_{\mu\nu}$ for the quasiparticles propagation in a (3+1)-dimensional relativistic, barotropic, irrotational fluid flow is given by
\begin{eqnarray}\label{ramnr}
g_{\mu\nu}=\frac{\rho}{\sqrt{1-u_{\alpha}u^{\alpha}/c_{0}^{2}}}\left[\eta_{\mu\nu}\left(1-\frac{u_{\alpha}u^{\alpha}}{c_{0}^{2}}\right)+\frac{u_{\mu}u_{\nu}}{c_{0}^{2}}\right]. 
\end{eqnarray}

Sometimes it is more convenient express the acoustic metric (\ref{ramnr}) as
\begin{equation}\label{ram}
g_{\mu\nu}=\rho\frac{{c}}{c_{s}}\left[\eta_{\mu\nu}+\left(1-\frac{c_{s}^{2}}{c^{2}}\right)\frac{v_{\mu}v_{\nu}}{c^{2}}\right],
\end{equation}
where $v^{\mu}=cu^{\mu}/||u||$ is the normalized four-velocity and the speed of sound $c_{s}$ is defined by
\begin{equation}\label{speedofsound}
c_{s}^{2}=\frac{c^{2}c_{0}^{2}/||u||^{2}}{1+c_{0}^{2}/||u||^{2}}.
\end{equation}

It is obvious from Eq. (\ref{ram}) that the acoustic metric $g_{\mu\nu}$ is disformally related to the background Minkowski spacetime. Writing in the lab coordinates ($x^{\mu}\equiv{c}t,x^{i}$), the relativistic acoustic line element takes the form
\begin{eqnarray}
ds^{2}=\rho\frac{{c}}{c_{s}}\left[\left(-1+\xi\frac{{v_0}^{2}}{c^{2}}\right)c^{2}dt^{2}+2\xi\frac{v_{0}v_{i}}{c^{2}}cdtdx^{i}+\left(\delta_{ij}+\xi\frac{v_{i}v_{j}}{c^{2}}\right)dx^{i}dx^{j}\right],
\end{eqnarray}
where $\xi\equiv\left(1-c_{s}^{2}/c^{2}\right)$. The normalization condition $v^{2}=-c^{2}$ allow us to rewrite the above acoustic line element as 
\begin{eqnarray}\label{relacoustmetric}
ds^{2}=\rho\frac{{c}}{c_{s}}\left[-\left(c_{s}^{2}-\xi\mathbf{v}^{2}\right)dt^{2}\pm2\xi\sqrt{1+\frac{\mathbf{v}^{2}}{c^{2}}}(v_{i}dx^{i})dt+\left(\delta_{ij}+\xi\frac{v_{i}v_{j}}{c^{2}}\right)dx^{i}dx^{j}\right],
\end{eqnarray}
where $\mathbf{v}^{2}=v_{i}v^{i}$ is the square normalized three-velocity. The acoustic metric in terms of the spatial three-velocity makes the interpretation more clear and allows easy comparison with Ref. \cite{Bilic:1999sq}.


\section{Analogue Black holes}\label{analogueBHs}

\subsection{Static acoustic spacetimes}

One of our purposes in this work is to demonstrate that static metric solutions can be incorporated in the formalism of relativistic acoustic spacetimes. Such a description can be achieved only when there is a Killing vector which is timelike at spatial infinity and hypersurface orthogonal. This means the existence of a time coordinate in which the metric can be written in diagonal form, i.e., one should be able to define a new time coordinate $\tau$ by the relation 
\begin{eqnarray}
cd\tau=cdt\mp\frac{\xi({v}_{0}/c^{2})v_{i}dx^{i}}{\left(-1+\xi{}v_{0}^{2}/c^{2}\right)},
\end{eqnarray}
where, for the moment, we are implicitly assuming that the vector $\xi({v}_{0}/c^{2})v_{i}/\left(-1+\xi{}v_{0}^{2}/c^{2}\right)$ is integrable, i.e., can be expressed as a gradient of some scalar. In terms of the new time coordinate $\tau$, the relativistic acoustic metric assumes the form
\begin{eqnarray}\label{staticacousticmetric}
ds^{2}=\rho\frac{c}{c_{s}}\left\lbrace-\left(c_{s}^{2}-\xi\mathbf{v}^{2}\right)d\tau^{2}+\left[\delta_{ij}+\frac{\xi{v_{i}v_{j}}}{\left(c_{s}^{2}-\xi\mathbf{v}^{2}\right)}\right]dx^{i}dx^{j}\right\rbrace,
\end{eqnarray}
which is explicitly static, rather than just stationary. The integrability constraint is satisfied if
\begin{eqnarray}
\epsilon_{ijk}\partial_{j}\left(\frac{\xi{v}_0v_{k}/c^{2}}{-1+\xi{}v_{0}^{2}/c^{2}}\right)=0.
\end{eqnarray}

Using the fact that the normalized four-velocity $v^{\mu}$ is hypersurface orthogonal, we obtain, after some computations, the condition
\begin{eqnarray}
\mathbf{v}\times\mathbf{\nabla}\left(c_{s}^{2}-\xi{\mathbf{v}^{2}}\right)=0.
\end{eqnarray}


\subsection{Horizons and ergosurfaces in relativistic acoustics}

Ergosurfaces and Killing horizons are important issues in general relativity. In the framework of nonrelativistic acoustic spacetimes, it is straightforward to define these notions \cite{Barcelo:2005fc,Visser:1993ub,Visser:1997ux}. Now, we want to see how these concepts arise in the relativistic extension. To begin with, let us for simplicity consider a stationary fluid. In this case, we can define a timelike Killing vector $\chi^{\mu}=\delta^{\mu}_{0}$, with its norm given by
\begin{eqnarray}\label{killingnorm}
\chi^{2}=g_{00}=-\rho\frac{c}{c_{s}}\left(c_{s}^{2}-\xi\mathbf{v}^{2}\right).
\end{eqnarray}

From the above relation (\ref{killingnorm}), it is clear that the condition $\mathbf{v}^{2}={c_{s}^{2}}/\xi$ defines an ergosurface. 
In a region where $\mathbf{v}^{2}>{c_{s}^{2}}/\xi$, the magnitude of the Killing vector field (\ref{killingnorm}) becomes spacelike, which characterizes an ergoregion in relativistic acoustic spacetimes. When the flow is static, the Killing horizon and the ergosurface coincide. 


\subsection{Surface gravity}

Surface gravity is important in both classical and semiclassical aspects of black hole physics. From an experimental point of view it is important to know what governs the surface gravity, as one wants to know how to maximize the Hawking temperature. Here we will be dealing only with static Killing horizons (though this is easily extendable to stationary spacetimes), so we will not concern ourselves with the complications of Ref. \cite{Cropp:2013zxi}, though such concerns will be equally relevant when considering non-Killing horizons in the context of relativistic fluids. 

Taking one of the standard definitions discussed in Ref. \cite{Cropp:2013sea} (in terms of the inaffinity of null geodesics), the surface gravity, $\kappa$ of the Killing horizon is given by
\begin{equation}
\nabla_\mu\left(\chi^2\right)=-2\kappa\chi_{\mu},
\end{equation}
From Eq. \eqref{killingnorm}, we can directly calculate this to be
\begin{equation}
\kappa=\frac{1}{2}\frac{d}{dr}\left(c_s^{2}-\xi\mathbf{v}^{2}\right), 
\end{equation}
where the speed of sound $c_{s}$ and the flow velocity $\mathbf{v}$ are taken at the Killing horizon. 


\subsection{Canonical acoustic black hole}\label{CAM}

Now, let us see what sort of static acoustic metric emerges when we take an incompressible spherically symmetric flow. Since the density $\rho$ is position independent, the continuity equation (\ref{cemrnc}) in spherical coordinates implies that $u^{r}=-c_{s}r_{0}^{2}/{r}^{2}$, where $r_{0}$ is a normalization constant and the minus sign indicates that we are considering an ingoing flow. One can also make the speed of sound $c_{s}$ a constant. According to definition (\ref{speedofsound}), the speed of sound $c_{s}$ depends on $c_{0}$, which is a function of the self-interaction potential $U$, given by the relation $c_{0}^{2}=(\hbar/2m)\rho{U''}$. In this sense, it is possible adjust the parameter $c_{0}$ (the potential $U$ to be more precise) in such a way that the speed of sound becomes position independent. In a laboratory setup, this configuration is possible by using the Feshbach resonance to control the interaction strength between the atoms \cite{Jain:2007gg,Weinfurtner:2007dq}. Furthermore, we remark that the hypersurface orthogonality (\ref{hypersurface}) imposes that the un-normalized four-velocity component $u^{0}$ cannot have any spacetime dependence in the static case. So, the relativistic acoustic metric (\ref{relacoustmetric}) in spherical coordinates is, up to a position-independent conformal factor,
\begin{eqnarray}\label{lecab}
ds^2&=&-\left[1-\xi\left(\frac{(c/u^{0})^{2}r_{0}^{4}}{r^{4}-(c_{s}/u^{0})^{2}r_{0}^{4}}\right)\right]c_{s}^{2}dt^2\pm\frac{2\xi(c/u^{0})r_{0}^{2}r^{2}}{r^{4}-(c_{s}/u^{0})^{2}r_{0}^{4}}c_{s}dtdr\nonumber\\
&&+\left[1+\xi\left(\frac{(c_{s}/u^{0})^{2}r_{0}^{4}}{r^{4}-(c_{s}/u^{0})^{2}r_{0}^{4}}\right)\right]dr^2+r^2d\Omega^2.
\end{eqnarray}
 
It is convenient to put the above acoustic metric (\ref{lecab}) in the diagonal form. Making the coordinate change 
\begin{equation}
dT=dt\pm\frac{\xi(u^{0}/c)r_{0}^{2}/r^2}{c_{s}\left[(u^{0}/c)^{2}-r_{0}^4/r^{4}\right]}dr, 
\end{equation}
the line element (\ref{lecab}) assumes the form
\begin{equation}\label{canonicalbh}
ds^2=-\left[1-\xi\left(\frac{(c/u^{0})^{2}r_{0}^{4}}{r^{4}-(c_{s}/u^{0})^{2}r_{0}^{4}}\right)\right]c_{s}^{2}dT^2+\frac{dr^2}{\left[1-\xi\left(\frac{(c/u^{0})^{2}r_{0}^{4}}{r^{4}-(c_{s}/u^{0})^{2}r_{0}^{4}}\right)\right]} +r^2d\Omega^2.
\end{equation}

Alternatively, one can express (\ref{canonicalbh}) as
\begin{equation}\label{alternativeformcanonicalbh}
ds^2=-\left(\frac{r^{4}-(c/u^{0})^{2}r_{0}^{4}}{r^{4}-(c_{s}/u^{0})^{2}r_{0}^{4}}\right)c_{s}^{2}dT^2+\left(\frac{r^{4}-(c_{s}/u^{0})^{2}r_{0}^{4}}{r^{4}-(c/u^{0})^{2}r_{0}^{4}}\right)dr^2+r^2d\Omega^2.
\end{equation}

From Eq.~(\ref{alternativeformcanonicalbh}) one immediately realizes that the Killing horizon ($\chi^{2}=0$) is located at
\begin{equation}
r_{\KH}=\sqrt{\frac{c}{u^{0}}}r_{0}.
\end{equation}

We also note that the relativistic acoustic line element (\ref{alternativeformcanonicalbh}) diverges at $r=\sqrt{\frac{c_{s}}{u^{0}}}r_{0}$. In fact, at this point the un-normalized rBEC four-velocity becomes null, i.e., $||u||^{2}=0$, and the normalized flow vector $v^{\mu}$ goes to infinity, meaning that the acoustic description breaks down. To conclude, we mention that the acoustic geometry (\ref{canonicalbh}) describes a spherically symmetric flow which has many of the same properties as a Schwarzschild black hole metric, but does not have counterpart in standard general relativity geometries. Because of that, we will refer to the above solution as the canonical acoustic black hole, in analogy with Ref. \cite{Visser:1997ux}. 


\subsection{Schwarzschild black hole}\label{schwarz}

Now we would like to know if it is possible to map the relativistic acoustic metric (\ref{relacoustmetric}) into the Schwarzschild metric. According to Eq~(\ref{staticacousticmetric}), the mapping between both metrics can be obtained only if the normalized radial flow assumes the form
\begin{equation}
\mathbf{v}^{2}=\frac{2GM}{\xi{r}}, 
\end{equation}
which can be done if the un-normalized radial flow is
\begin{eqnarray}
u^{r}=u^{0}\sqrt{\left(\frac{2GM}{2GM+\xi{c^{2}}r}\right)}. 
\end{eqnarray}

Nevertheless, the above constraint does not satisfy the continuity equation for an incompressible fluid. To overcome this difficulty, we consider a fluid with a nonconstant density $\rho$. In this case, the (3+1)-dimensional continuity equation imposes that 
\begin{eqnarray}
\rho=\frac{A}{u^{0}r^{2}}\sqrt{\left(1+\frac{\xi{c^{2}}r}{2M}\right)},
\end{eqnarray}
where $A$ is a position-independent factor. Indeed, as in the case of nonrelativistic acoustic geometries, the Schwarzschild metric can be mimicked at most up to a (nonconstant) conformal factor, given by
\begin{eqnarray}
ds^{2}\propto\left\lbrace-\left(1-\frac{2GM}{rc_{s}^{2}}\right)c_{s}^{2}d\tau^{2}+\left(1-\frac{2GM}{rc_{s}^{2}}\right)^{-1}dr^{2}+r^{2}d\Omega^{2}\right\rbrace,
\end{eqnarray}
where we are picking the speed of sound to be a position independent constant.

For many purposes this will be enough as many features (causal features and surface gravity, for example) are conformally invariant.  


\subsection{Nonrelativistic limit}

We have shown in the previous sections that some concepts that are very useful in general relativity can be easily identified in the framework of the relativistic acoustic spacetimes. Now, in order to check that the formalism is fully consistent, one needs to ensure that the nonrelativistic limit can be obtained in a smooth way. To start, we note that there are many considerations to take into account, which we shall outline. The conditions on the fluid in which such behavior can be achieved was fully discussed in Ref. \cite{Fagnocchi:2010sn}. First, to reproduce the nonrelativistic regime, the self-interaction between the atoms must be weak, i.e. $c_{0}\ll{c}$. Also, in this regime, $u^{0}\rightarrow{c}$. In addition, the speed of sound (\ref{speedofsound}) reduces to $c_{0}$. It is important to note that in the nonrelativistic regime the flow velocity $v^{i}$ and the speed of sound $c_{s}$ are much smaller than the speed of light $c$. With these conditions, the normalized four-velocity assumes the form $v^{\mu}\approx(c;u^{i})$. Thus, under these assumptions, we also have
\begin{eqnarray}
\xi=1+\frac{c_{s}^{2}}{c^{2}}\approx1, \quad\quad\quad 1+\frac{\mathbf{v}^{2}}{c^{2}}\approx1,
\end{eqnarray}
which implies that the relativistic acoustic line (\ref{relacoustmetric}) assumes the form
\begin{eqnarray}\label{nonrelativisticacousticmetric}
ds^{2}=\frac{\rho_{m}}{c_{s}}\left[-\left(c_{s}^{2}-{\mathbf{u}^{2}}\right)dt^{2}\pm2u_{i}dx^{i}dt+\delta_{ij}dx^{i}dx^{j}\right],
\end{eqnarray}
where $\rho_{m}$ is the mass density and $\mathbf{u}^{2}=u_{i}u^{i}$ is the square un-normalized three-velocity. The line element (\ref{nonrelativisticacousticmetric}) is the standard nonrelativistic acoustic metric. 

We remark that in \cite{Bilic:1999sq} $u^0$ was always taken to be $c$, simplifying many expressions down to their familiar forms. However, in \cite{Bilic:1999sq} the author considered a perfect fluid scenario, for which there is no scale at which the analogue metric breaks down. In our case, while taking the nonrelativistic limit, we must ensure that we remain within a low-energy regime where the higher-order terms in the dispersion are still small: this depends on conditions on $u^0$, which we therefore need to keep general. Finally, let us stress that when considering the above assumptions on the rBEC, all the previous definitions of acoustic quantities reduce to the standard nonrelativistic case.


\section{Einstein--{\AE}ther black holes and universal horizons}\label{universal}

Einstein--{\AEther} and \Horava--Lifshitz gravity are two closely related Lorentz-violating theories of gravity that are diffeomorphism invariant but violate local Lorentz invariance. In the case of Einstein--{\AEther} Lorentz invariance is violated by introducing a preferred observer $a^{\mu}$, while \Horava--Lifshitz gravity introduces a preferred foliation defined by a scalar field $\Gamma$ known as the khronon. 

\subsection{Einstein--{\AEther} theory}

Einstein--{\AE}ther theory, (for general background see Refs.~\cite{Jacobson:2000xp, Jacobson:2010mx, Gasperini:1987nq, Gasperini:1998eb, Jacobson:2008aj, Eling:2004dk}), is a Lorentz-violating theory which still maintains many of the nice features of general relativity, such as general covariance and second-order field equations. This is done through introducing a timelike unit vector field, $a^{\mu}$, known as the {\ae}ther. The action is given by

\begin{equation}
S=\frac{1}{16\pi G}\int \d^4 x\sqrt{-g} (R+\mathcal{L}_{ae})\,; \qquad \mathcal{L}_{ae}=-Z^{\mu\nu}{}_{\alpha\beta}\,(\nabla_{\mu}a^{\alpha})(\nabla_{\nu}a^{\beta})+\lambda(a^2+1).
\label{ac:ae}
\end{equation}
Here $\lambda$ is a Lagrange multiplier, enforcing the unit timelike constraint on $a^{\mu}$, and $Z^{\mu\nu}{}_{\alpha\beta}$ couples the \aether\ to the metric through four distinct coupling constants:
\begin{equation}
Z^{\mu\nu}{}_{\alpha\beta}=c_1g^{\mu\nu}g_{\alpha\beta}+c_2\delta^{\mu}{}_{\alpha}\delta^{\nu}{}_{\beta}+c_3\delta^{\mu}{}_{\beta}\delta^{\nu}{}_{\alpha}-c_4 a^{\alpha}a^{\nu}g_{\alpha\beta}.
\end{equation}
It is often useful to work with combinations of these constants for which we will adopt the convenient notation whereby $c_{14}=c_1+c_4$, \emph{etc}. 

\subsection{\Horava--Lifshitz gravity}

\Horava--Lifshitz (HL) gravity (see for example Ref.~\cite{Sotiriou:2010wn} for a review) was motivated by the possibility of achieving renormalizability by adding to the action the terms containing higher-order spatial derivatives of the metric, but no higher-order time derivatives, so as to preserve unitarity.   This procedure naturally leads to a foliation of spacetime into spacelike hypersurfaces.

Power-counting renormalizability requires the action to include terms with at least six spatial derivatives in four dimensions~\cite{Horava:2008ih,Horava:2009uw,Visser:2009fg}, but all lower-order operators compatible with the symmetry of the theory are expected to be generated by radiative corrections, so the most general action takes the form~\cite{Blas:2009qj}
\begin{equation}
\label{SBPSHfull}
S_{HL}= \frac{M_{\rm Pl}^{2}}{2}\int \d t\, \d^3x \, N\sqrt{h}\left(L_2+\frac{1}{M_\star^2}\;L_4+\frac{1}{M_\star^4}\;L_6\right)\,,
\end{equation}
where $h$ is the determinant of the induced metric $h_{ij}$ on the spacelike hypersurfaces, while
\begin{equation}
L_2=K_{ij}\,K^{ij} - \lambda K^2 
+ \xi\, {}^{(3)}\!R + \eta b_ib^i\,,
\end{equation}
where $K$ the trace of the extrinsic curvature $K_{ij}$, ${}^{(3)}\!R$ is the Ricci scalar of $h_{ij}$, $N$ is the lapse function, and $b_i=\partial_i \ln N$. The quantities $L_4$ and $L_6$ denote a collection of fourth- and sixth-order operators respectively, and $M_\star$ is the scale that suppresses these operators (which does not coincide {\em a priori} with $M_{\rm Pl}$). 

In Ref.~\cite{Jacobson:2010mx} (see also Ref.~\cite{Afshordi:2009tt}) it was shown that  the solutions of  Einstein--\Aether\ theory are also the solutions of the infrared limit of \Horava--Lifshitz gravity if the \aether\ vector is assumed to be hypersurface orthogonal before the variation. More precisely, hypersurface orthogonality can be imposed through the local condition
\begin{equation}
a_\mu=\frac{\partial_\mu \Gamma} {\sqrt{-g^{\alpha\beta} \;\partial_\alpha \Gamma \, \partial_\beta \Gamma}}\, ,
\label{eq:orthogonal}
\end{equation}
where $\Gamma$ is a scalar field that defines a foliation (often named for this reason the ``khronon"). Choosing $\Gamma$ as the time coordinate one selects the preferred foliation of HL gravity, and the action (\ref{ac:ae}) reduces to the action of the infrared limit of \Horava--Lifshitz gravity, whose Lagrangian we denoted as $L_2$ in Eq.~(\ref{SBPSHfull}). The details of the equivalence of the equations of motion and the correspondence of the parameters of the two theories can be found in Refs.~[\cite{Barausse:2012ny, Barausse:2012qh, Jacobson:2013xta}].

This fact is particularly relevant for the present investigation. Indeed we shall consider here static, spherically symmetric black hole solutions in Einstein--{\AEther} for which the \aether\ field is always hypersurface orthogonal. Hence such solutions of Einstein--{\AEther} are also solutions of \Horava--Lifshitz gravity (at least in the infrared limit when one neglects the $L_4$ and $L_6$ contributions to the total Lagrangian). We shall therefore consider, from here on, black hole solutions in Einstein--{\AEther} theory.

\subsection{{\AE} black holes}

Black holes in Einstein--{\AE}ther theory have been extensively considered in recent years (see for example Refs.~\cite{Barausse:2011pu, Blas:2011ni, Berglund:2012bu, Berglund:2012fk, Eling:2006ec, Mohd:2013zca}). Among the most striking results concerning these solutions was the realization---in the (static and spherically-symmetric) black hole solutions of both Einstein--{\AE}ther and \Horava--Lifshitz gravity---that they seem generically to be endowed with a new structure that was soon christened the universal horizon~\cite{Barausse:2011pu, Blas:2011ni}. 

These universal horizons can be described as compact surfaces of constant khronon field and radius while nothing singular happens to the metric. Given that the khronon field defines an absolute time, any object crossing this surface from the interior would necessarily also move back in absolute time (the {\ae}ther time), something forbidden by the definition of causality  in the theory. Another way of saying this is that even a particle capable of instantaneous propagation---light cones opened up to an apex angle of a full 180 degrees, something in principle possible in Lorentz-violating theories---would just move around on this compact surface and hence be unable to escape to infinity. Hence the name universal horizon; even the superluminal particles would not be able to escape from the region it bounds. 


\subsection{Exact solutions}
For some specific combinations of the coefficients, $c_i$ there are explicit, exact black hole solutions. In particular, two exact solutions for static, spherically symmetric black holes have been found. As we will use these solutions extensively throughout this paper, we will briefly summarize some of their relevant details. For more information and background we refer the reader to Ref.~\cite{Berglund:2012bu}. Both solutions, in Eddington--Finkelstein coordinates, can be written as 
\begin{equation}
\d s^2 =-e(r)\;\d v^2 +2\,\d v\,\d r +r^2 \; \d \Omega^2.
\end{equation}
Here the form of the \aether\ is
\begin{equation}
a^a=\left\lbrace \alpha(r), \beta(r), 0, 0 \right\rbrace; \qquad a_a=\left\lbrace \beta(r)-e(r)\alpha(r), \alpha(r), 0, 0\right\rbrace.
\end{equation}
Note that from the normalization condition, $u^2=-1$, there is a relation between $\alpha(r)$ and $\beta(r)$:
\begin{equation}
\beta(r)=\frac{e(r)\alpha(r)^2-1}{2\alpha(r)}.
\end{equation}
The two known exact black hole solutions to Einstein--{\AEther} theory correspond to the special combinations of coefficients $c_{123} = 0$ and $c_{14}=0$. For instance, in the $c_{123} = 0$ case the metric and \aether\ take the form
\begin{equation}\label{c123solution}
e(r)=1-\frac{r_0}{r}-\frac{r_u(r_0+r_u)}{r^2}; \qquad\hbox{where}\qquad  r_u = \left[\sqrt{\frac{2 - c_{14}}{2(1 - c_{13})}} - 1\right]\frac{r_0}{2}.
\end{equation}
Here is $r_0$ is essentially the mass parameter. Furthermore
\begin{equation}
\alpha(r)=\left(1+\frac{r_u}{r}\right)^{-1}; \qquad \beta(r)=-\frac{r_0+2r_u}{2r} \qquad \chi \cdot u=-1+\frac{r_0}{2r}.
\end{equation}
For this particular exact solution, the Killing horizon is located at $r_{\KH}=r_0+r_u$, and the universal horizon is at $r_{\UH}={r_0}/{2}$. Note that this includes a possible $r_u=0$ case, where the metric assumes the same form as the Schwarzschild black hole, but with the presence of the additional \aether\ field. 


\section{Analogue universal horizons}\label{analogueuh}

\subsection{Acoustic {\ae}ther in analogue gravity}

As we have discussed in the Introduction, universal horizons cannot be simulated in the formalism of analogue gravity using just the acoustic metric. The reason is due to a lack of freedom in how we pick the \aether. To overcome this issue, one needs add more degrees of freedom in the system. We proposed to incorporate the \aether\ in the following way: first, one needs to consider a modification of the nonlinear Klein--Gordon equation (\ref{nlkg}) through the introduction of an external field $\Phi$ (here, $\Phi$ represents an arbitrary field which is not restricted to be a scalar; in fact, we will consider interactions with four-vector fields), i.e., $V\ne0$ in Eq.~(\ref{nlkg}). Then, performing the linearization, and under the standard assumptions in order to have the acoustic description of the phonons, the equation for the quantum fluctuation $\psi$ will be split into terms containing the metric plus a potential, i.e.,
\begin{equation}\label{modifiedequation}
{\Large\Box}\psi+V_{\rm eff}\psi=0, 
\end{equation}
where 
\begin{equation}\label{acoustickg}
{\Large\Box}=\frac{1}{\sqrt{-g}}\partial_{\mu}\left(\sqrt{-g}g^{\mu\nu}\partial_{\nu}\right). 
\end{equation}

In Eq.~\eqref{acoustickg}, the acoustic metric $g_{\mu\nu}$ will be defined by the uncoupled fluid flow $u$ while the emergent potential $V_{\rm eff}$ will be a function of the external field $\Phi$. The presence of this new term in the description of the quasiparticles induces the appearance of a modified dispersion relation, which characterizes the presence of supersonic modes in the rBEC. In \AE\ and HL gravity theories, the higher-order momenta terms in the dispersion relation for matter fields are assumed to be induced by the coupling of the matter fields with the \aether. In this way, taking into account that in Eq.~(\ref{modifiedequation}) the dispersion relation of the quasiparticles is changed by the presence of the external field, it becomes natural to relate the \aether\ to the field $\Phi$. Applying the normalization condition $a^{2}=-c_{s}^{2}$, we ensure that the acoustic \aether\ is timelike everywhere in the acoustic spacetime. To be consistent, the acoustic \aether\ needs to be aligned to the hypersurface orthogonal timelike Killing vector at infinity. Also, since the universal horizon is a spacelike hypersurface, it must be inside the Killing horizon.
\\

Following the above proposal, a straightforward way to couple an external field to the rBEC would be through the electromagnetic minimal prescription, which is given by 
\begin{eqnarray}\label{emmc1}
\partial_{\mu}&\rightarrow&D_{\mu}=\partial_{\mu}+\frac{iq}{c\hbar}A_{\mu},
\end{eqnarray}
where $q$ is the coupling constant and $A^{\mu}$ is the gauge field. The $U(1)$ gauge-invariant Lagrangian describing the interaction of the complex scalar field $\phi$ with the gauge field $A^{\mu}$ is defined by
\begin{eqnarray}\label{gaugelagrangian}
\mathcal{L}=-\eta^{\mu\nu}\left(D_{\mu}\phi\right)^{*}\left(D_{\nu}\phi\right)-m^{2}\phi^{*}\phi-\lambda(\phi^{*}\phi)^{2}-\frac{1}{4}F_{\mu\nu}F^{\mu\nu},
\end{eqnarray}
where $F_{\mu\nu}(=\partial_{\mu}A_{\nu}-\partial_{\nu}A_{\mu})$ is the field strength and $m$ and $\lambda$ are parameters that under subtle conditions can trigger a spontaneous symmetry breaking and form a charged condensate. After the condensation (for further details, see Ref. \cite{Kapuska:1981jj}), the charged rBEC is described by the gauge-invariant equation
\begin{equation}
\left[\Box-m^{2}+2\frac{iq}{c\hbar}A^{\mu}\partial_{\mu}+\frac{iq}{c\hbar}\partial_{\mu}A^{\mu}-\left(\frac{q}{c\hbar}\right)^{2}A_{\mu}A^{\mu}-U'(\rho;\lambda)\right]\phi=0, 
\end{equation}
where $U\equiv\lambda(\phi^{*}\phi)^{2}$ and $U'=dU/d\phi$. At the linearized perturbations level, and under suitable assumptions, the charged phonons propagate under an effective acoustic metric \cite{Cropp:2015tua}. Nevertheless, following our prescription and relating the acoustic metric to the fluid flow $u^{\mu}$ and the \aether\ to the gauge field $A^{\mu}$, will lead us to an inconsistent scenario where the properties of the \aether\ (including the existence of possible universal horizons) are mere gauge effects. 

For the sake of clarity, let us consider the continuity equation in the Madelung representation, namely, 
\begin{eqnarray}\label{gaugecontinuityequation}
\partial_{\mu}j^{\mu}=\partial_{\mu}\left[\rho\left(u^{\mu}+\frac{q}{mc}A^{\mu}\right)\right].
\end{eqnarray}
According to Eq.~(\ref{gaugecontinuityequation}), the net effect of the electromagnetic minimal coupling is just a shift in the four-velocity of the condensate. But, since the four-current is gauge invariant, one cannot uncouple the gauge field $A^{\mu}$ and the four-vector $u^{\mu}$. In other words, the gauge symmetry does not allow us to split the scalar and gauge fields. It means that building the acoustic metric with the four-vector $u^{\mu}$ and the \aether~with the gauge field $A^{\mu}$ will lead to gauge-dependent acoustic quantities. Other gauge couplings would suffer from the same disease. We note, however, that although gauge couplings cannot be used to simulate analogue universal horizons, they provide a natural way to incorporate vorticity in analogue systems \cite{Cropp:2015tua}. 
\\

To consistently and unambiguously model the metric and \aether\ let us suppose that our system is described by the following Lagrangian density 
\begin{eqnarray}
\mathcal{L}=-\eta^{\mu\nu}\left(D_{\mu}\phi\right)^{*}\left(D_{\nu}\phi\right)-m^{2}\phi^{*}\phi-\lambda(\phi^{*}\phi)^{2}-\frac{1}{4}F_{\mu\nu}F^{\mu\nu}-\frac{1}{2}m_{A}^{2}A_{\mu}^{2},
\end{eqnarray}
where the only difference with the Lagrangian (\ref{gaugelagrangian}) is the presence of a mass term $\frac{1}{2} m_{A}^{2}A_{\mu}^{2}$, which explicitly violates the gauge symmetry. Such massive electromagnetic fields are known as Proca fields \cite{martinshaw}. In this framework, the equations of motion for the scalar and vector fields are given by
\begin{eqnarray}\label{Procafieldequations}
\left[\Box-m^{2}+2\frac{iq}{c\hbar}A^{\mu}\partial_{\mu}+\frac{iq}{c\hbar}\partial_{\mu}A^{\mu}-\left(\frac{q}{c\hbar}\right)^{2}A_{\mu}A^{\mu}-U'(\rho;\lambda)\right]\phi&=&0, \\\label{procafe}
\partial_{\mu}F^{\mu\nu}-m_{A}^{2}A^{\nu}&=&-j^{\nu},
\end{eqnarray}
where the associated conserved current $j^{\mu}$ is
\begin{eqnarray}\label{ccproca}
j^{\mu}=i\left[\phi(D^{\mu}\phi)^{*}-\phi^{*}(D^{\mu}\phi)\right]. 
\end{eqnarray}

It follows from the Eq.~(\ref{procafe}) that if the source current is conserved ($\partial_{\mu}j^{\mu}=0$), or if there are no sources ($j^{\mu}=0$), then
\begin{eqnarray}\label{lorenzcondition}
\partial_{\mu}A^{\mu}=0
\end{eqnarray}
for $m_{A}\ne0$, which is a constraint in the Proca electromagnetism called the Lorenz condition.

Now, decomposing the complex scalar field $\phi$ into an amplitude $\rho$ and a phase $\theta$ through the Madelung representation, the conserved current (\ref{ccproca}) takes the same form as Eq.~(\ref{gaugecontinuityequation}).

Proceeding exactly as in the rBEC to find the equation for the perturbations, i.e., inserting ${\phi}=\varphi(1+{\psi})$ into Eq.~(\ref{Procafieldequations}), we get to the linearized fluctuations
\begin{eqnarray}\label{gcrbec}
\left\lbrace\left[i\hbar{u^{\mu}}\partial_{\mu}+i\hbar\frac{q}{mc}{A^{\mu}}\partial_{\mu}+T_{\rho}\right]\frac{1}{c_{0}^{2}}\left[-i\hbar{u^{\nu}}\partial_{\nu}-i\hbar\frac{q}{mc}{A^{\nu}}\partial_{\nu}+T_{\rho}\right]-\frac{\hbar^{2}}{\rho}\eta^{\mu\nu}\partial_{\mu}\rho\partial_{\nu}\right\rbrace\psi=0.
\end{eqnarray}

Making use of the continuity equation and neglecting the quantum potential $T_{\rho}$ under the same assumptions as in Ref. \cite{Cropp:2015tua}, Eq.~(\ref{gcrbec}) takes the form
\begin{eqnarray}\label{acousticequation}
&&\partial_{\mu}\left[\frac{\rho}{c_{0}^{2}}{u^{\mu}}{u^{\nu}}-\rho\eta^{\mu\nu}+\frac{\rho}{c_{0}^{2}}\frac{q}{mc}\left(u^{\mu}A^{\nu}+A^{\mu}u^{\nu}\right)+\frac{\rho}{c_{0}^{2}}\frac{q^{2}}{m^{2}c^{2}}A^{\mu}A^{\nu}\right]\partial_{\nu}\psi=0,
\end{eqnarray}
which shows that the massless quasiparticles propagate under an effective acoustic geometry. It is important to note that Eq.~(\ref{gcrbec}) leads to a dispersion relation similar to that found in Ref. \cite{Fagnocchi:2010sn}. The difference lies in that, in our system, the coupling with the vector field $A^{\mu}$ causes a shift in the four-velocity $u^{\mu}\rightarrow{f^{\mu}\equiv u^{\mu}+(q/mc)A^{\mu}}$. When taking into account the massless modes in the low-momentum limit, the dispersion relation of Eq.~(\ref{gcrbec}) takes the form
\begin{eqnarray}
w^{2}\approx{c^{2}}\left[\frac{(c_{0}/f^{0})^{2}k^2}{1+(c_{0}/f^{0})^{2}}+\frac{k^{4}}{4(mf^{0}/\hbar)^{2}(1+(c_{0}/f^{0})^{2})^{3}}\right].
\label{disprel}
\end{eqnarray}
In the UV regime of the above dispersion relation, i.e., $k^{4}\gg{k}^{2}$, the massless modes become supersonic. 

According to Eq.~(\ref{acousticequation}), one can proceed as usual in the analogue gravity formalism and determine the acoustic metric using the flow vector $u^{\mu}$ and the vector field $A^{\mu}$. Alternatively, one can also split Eq.~(\ref{acousticequation}) in such a way that the acoustic metric is defined only by the uncoupled fluid $u^{\mu}$ and the remaining terms are associated to an effective potential, i.e.,
\begin{eqnarray}\label{kgerac}
\partial_{\mu}\left[\frac{\rho}{c_{0}^{2}}{u^{\mu}}{u^{\nu}}-\rho\eta^{\mu\nu}\right]\partial_{\nu}\psi+\,\partial_{\mu}\left[\frac{\rho}{c_{0}^{2}}\frac{q}{mc}\left(u^{\mu}A^{\nu}+A^{\mu}u^{\nu}\right)+\frac{\rho}{c_{0}^{2}}\frac{q^{2}}{m^{2}c^{2}}A^{\mu}A^{\nu}\right]\partial_{\nu}\psi=0. 
\end{eqnarray}

This ambiguity in how to describe the modes also appears in Lorentz-violating gravity theories. Generically, the matter fields couple to the \aether, which leads to an ultraviolet modified dispersion relation. These modified dispersion relations arise from a modified field equation for the particles by adding higher-order kinetic terms to the standard d'Alembertian. In the regime of low energy, i.e., neglecting the higher-order operators in the equation for such modes, the equation reduces to the usual Klein-Gordon equation. In this limit, each of these particles will travel under an effective metric disformally related to the original one and given by the metric redefinition
\begin{equation}
g^{(i)}_{\mu\nu}=g_{\mu\nu}+\left(\sigma^{2}_{i}-1\right)a_{\mu}a_{\nu},
\end{equation}
where $\sigma^{2}_{i}$ is the propagation speed of the $i$th spin mode. When considering only second-order field equations, it is merely a matter of convenience to adopt either metric definition to describe the particle propagation in these theories since both are physically equivalent. In our prescription, a similar situation happens. Of course in the presence of several fields this ambiguity is somewhat resolved in favor of the standard metric $g_{\mu\nu}$ as this will be the only universal structure in the game. In our present model the phononic excitations are unique (there is a high-energy gapped mode as well which does not share the same speed of sound of the massless one) although an analogue of a multifield situation could be realized in the case of a relativistic generalization of a multicomponent BEC background (see e.g. Ref.~\cite{Liberati:2005pr}). Nonetheless, the natures of the metric and \aether\ contributions are clearly rather different; the first is generated by the BEC flow while the latter is due to an superimposed (and {\em a priori} independent) external field. For all these reasons we shall treat the role of the external field separately from that of the flow-induced metric.

So, assuming the acoustic metric is described by the uncoupled fluid $u^{\mu}$, i.e.~the standard relativistic acoustic metric (\ref{relacoustmetric}) for the phonons, and the remaining terms to an effective potential, one can associate the massive vector field $A^{\mu}$ to the \aether\ by setting
\begin{eqnarray}\label{acousticaether}
a^{\mu}=c_{s}\frac{A^{\mu}}{\sqrt{-g_{\alpha\beta}A^{\alpha}A^{\beta}}}, 
\end{eqnarray}
where $a^{\mu}$ is normalized according to emergent acoustic metric and satisfies the normalization condition $a^{2}=-c_{s}^{2}$, which ensures that the \aether\ is globally timelike in the acoustic geometry.

The definition (\ref{acousticaether}), however, needs to be dealt with carefully. In fact, as we have already mentioned, in \Horava--Lifshitz gravity the \aether\ must be always hypersurface orthogonal, while in the Einstein--\Aether\ theory the hypersurface orthogonality is required only in the static spherically symmetric spacetime. But, according to definition (\ref{acousticaether}), the hypersurface orthogonality condition is not satisfied, which can be seen explicitly by the relation
\begin{eqnarray}\label{hypersurfacecondition}
a_{[\mu}\nabla_{\nu}a_{\alpha]}&=&-\frac{1/6}{g_{\beta\gamma}A^{\beta}A^{\gamma}}\left[A_{\mu}F_{\nu\alpha}+A_{\nu}F_{\alpha\mu}+A_{\alpha}F_{\mu\nu}\right]. 
\end{eqnarray}
Consequently, the model under investigation is more closely related to the Einstein--\Aether\ gravity than \Horava--Lifshitz theory. However, when we impose spherical symmetry, the hypersurface condition (\ref{hypersurfacecondition}) is fully satisfied, i.e., $a_{[\mu}\nabla_{\nu}a_{\alpha]}=0$. 

To conclude this section, we remark that within this proposal, nonrelativistic BECs cannot accommodate an acoustic \aether, due to the fact that in the hydrodynamical description, potential terms do not give contributions upon linearization to the quasiparticles' equation, which means that we do not have a four-vector object at hand to define the acoustic \aether.


\subsection{Analogue Einstein--{\AE}ther black hole}

The coupling of the scalar field with the massive four-vector $A^{\mu}$ in the previous section provided a mechanism to introduce the {\aether} field in the context of analogue gravity. Therefore, we are now ready to investigate the existence of universal horizons in acoustic spacetimes. With respect to the aforementioned black hole solution for definiteness we will briefly consider the $c_{123}$ solution. To begin with, one can see from Eq.~(\ref{kgerac}) that a position-dependent density $\rho$  introduces a conformal interaction in the description of the quasiparticles. This occurs because the density $\rho$ is related to the conformal factor, which can be explicitly seen in Eq.~(\ref{relacoustmetric}). However, conformal transformations are not a symmetry of the Einstein--\Aether\ theory. In addition, the universal horizon is not conformally invariant. To circumvent this issue, the density $\rho$ needs to be position independent. This choice does not allow us to consider the Schwarzschild acoustic metric or any of the known exact solutions for Einstein--\Aether\ gravity. Nevertheless, we still can make use of the canonical acoustic black hole solutions found in Sec. \ref{CAM}. Since this class of solutions are exact, i.e., they do not depend on a position-dependent conformal factor, the issue related to the conformal interactions discussed above does not concern us. Besides that, one immediately realizes that by considering a position-independent density $\rho$, together with the Lorenz condition (\ref{lorenzcondition}), the continuity equation (\ref{gaugecontinuityequation}) reduces to 
\begin{equation}
\partial_{\mu}u^{\mu}=0, 
\end{equation}
reproducing solutions of the uncoupled continuity equation (\ref{cemrnc}) when the density $\rho$ is a constant. In this way, one can proceed through the use of the canonical black hole solution, where the un-normalized four-velocity $u^{\mu}$ is given by
\begin{equation}\label{unnormalizedfourvelocity}
u^{\mu}=\left(u^{0},\, -c_{s}r_{0}^{2}/r^{2}\right), 
\end{equation}
and the acoustic line element assumes the form (\ref{lecab}). In addition, as discussed in Sec. \ref{CAM}, the un-normalized four-velocity component $u^{0}$ and the speed of sound $c_{s}$ do not have spacetime dependence.

Since we have defined the acoustic metric that will be used to find the universal horizon, let us see the explicit form of the acoustic aether. The definition (\ref{acousticaether}) requires, besides the acoustic metric (\ref{lecab}), the specific form of the Proca field $A^{\mu}$. In order to find the solutions of such massive fields, one needs to solve the equations of motion (\ref{procafe}) and the Lorenz condition (\ref{lorenzcondition}) in the static regime. Taking into account the previous assumptions on the conserved current $j^{\mu}=\rho\left({u^{0}};u^{i}=-c_{s}r_{0}^{2}/r^{2}\right)$, one promptly finds
\begin{eqnarray}\label{Procasolutions}
A^{\mu}=\left\lbrace{A}^{T}=\left({B}\frac{e^{-m_{A}r}}{r}-\frac{c\rho}{m_{A}^{2}}\right), \,A^{r}=\left(-\frac{R_{0}^{2}}{2r^{2}}\right),A^{\theta}=A^{\phi}=0\right\rbrace, 
\end{eqnarray}
where $B$ is an integration constant, $A^{0}$ comes from Eq.~(\ref{procafe}) and $A^{r}$ is obtained from the Lorenz condition (\ref{lorenzcondition}). The solutions (\ref{Procasolutions}), together with the canonical acoustic BH metric (\ref{lecab}), determine the form of the acoustic \aether\ (\ref{acousticaether}). 

\subsection{Acoustic universal horizon}

Now that we have found an explicit solution for the acoustic {\aether}, we need to verify that our system can really mimic the universal horizon. To begin with, we note that $\chi\cdot{a}=a_{0}$, where $a_{0}$ is a function of $r$. The universal horizon is located at the point where the function $a_{0}=0$. Moreover, since the universal horizon is a spacelike hypersurface, it must lie inside the Killing horizon. 

Since the $a_{0}$ function depends on the fluid variables and the massive vector field, there are many possible parameter choices that can simulate the universal horizon in our formalism. To be consistent, all these multiple choices are required to ensure the existence of the universal horizon ($a_{0}=0$) and that this special hypersurface be located inside the Killing horizon ($r_{UH}<{r_{KH}}$). Figure \ref{fig1} below shows the general behavior for some arbitrary constants.

\begin{figure}[h]
\begin{center}
\includegraphics[scale=1.2]{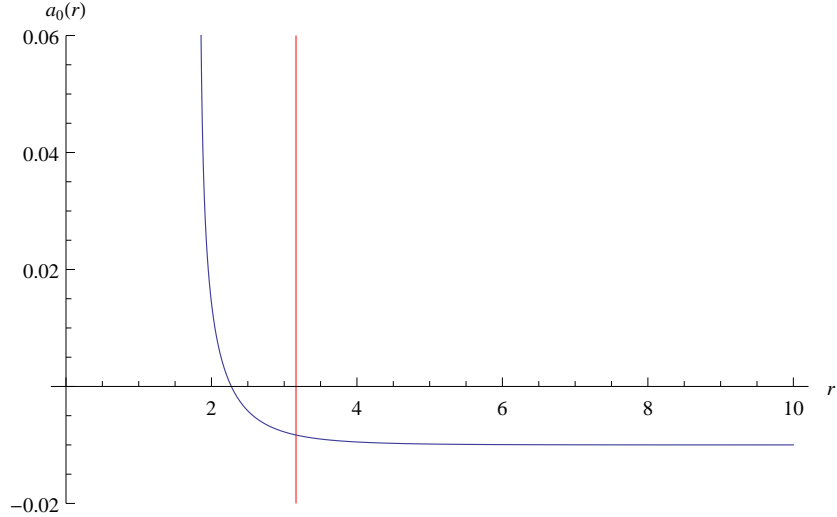}
\end{center}
\vspace{-0.5cm}
\caption{The $a_{0}$ component of the {\ae}ther field in terms of the radial parameter $r$. The point $r=2.26$ is the universal horizon, which is inside the Killing horizon (brown line). In our computations, we have used $B=1$ and $m=0.1$ for Eq.~(\ref{Procafieldequations}), $R_{0}=1$ in Eq.~(\ref{unnormalizedfourvelocity}) and the speed of sound $c_{s}=0.1c$.}\label{fig1}
\end{figure}

To conclude this subsection, we remark that the quantity $g_{\mu\nu}A^{\mu}A^{\nu}$ can be zero. Since the vector field $A^{\mu}$ is well behaved in the regime of validity of the acoustic description, Eq.~(\ref{acousticaether}) implies that the acoustic \aether\ must diverge when $g_{\mu\nu}A^{\mu}A^{\nu}=0$. This basically reflects the fact that in lab coordinates the definition (\ref{acousticaether}) is not well defined everywhere in the acoustic geometry. In principle, this divergence may arise outside the Killing horizon, which would be catastrophic to our model. However, there is a broad class of parameters for which this divergence is found inside the universal horizon, which ensures the consistency of our model.

\subsubsection{Ray tracing}

The universal horizon can be seen to be the true casual barrier by studying the behavior of rays, as in Ref. \cite{Cropp:2013sea}. Consider the dispersion relation \eqref{disprel}, given with respect to the lab time. Here, we are going to need to consider the dispersion relation in both the lab frame and the \aether\ frame, and relate this to the conserved energy along the ray, $\Omega \equiv g_{ab}\chi^ak^b$. For later convenience, we also define $k_r=k_a \rho^a$, for $\rho$ orthogonal to and the same magnitude as, $\chi$. The key point will be to show once there is a superluminal dispersion relation in \emph{some} frame, the existence of \emph{one} physically relevant frame where $\chi\cdot a=0$ leads to this surface being a causal barrier.  

We define the normalized lab frame $\left\{t, x\right\}$ such that $k_at^a= \omega, k_ax^a=k$, and we recover Eq.~\eqref{disprel}. This is fundamentally the local frame in which the dispersion relation is defined, and which we could work with in the absence of the \aether\ coupling. With the coupling to the \aether, we may also choose to work in the \aether\ (orthonormal) frame, $\left\{a, s\right\}$, which does not have to align with the $\left\{t, x\right\}$ frame. This is somewhat different from the Einsein-\Aether\ case, where naturally the dispersion relation is defined in the \aether\ frame.  

In the \aether\ frame we can define  $k_aa^a= \omega_a, k_as^a=k_s$, which are physically relevant quantities, but for which we do not, {\it a priori}, have an explicit relation $\omega_a(k_s)$. Note that at infinity automatically $\Omega=\omega_a=\omega$, $k=k_s=k_r$. For low-momentum, in the $\left\{t, x\right\}$ frame, we can simplify Eq.~\eqref{disprel} to
\begin{equation}
\omega=b_1k+b_3k^3,
\end{equation}
where
\begin{equation}
b_1=  \frac{c_0/f^0}{\sqrt{1+(c_0/f^0)^2}}; \qquad b_3 = \frac{\sqrt{1+(c_0/f^0)^2}}{8(mf^0/\hbar)(c_0/f^0)^2\left[1+ (c_0/f^0)^2\right]}.
\end{equation}
We can now decompose the conserved energy in two different ways
\begin{equation}
\Omega= -\omega_a (a\cdot \chi) \pm k_s (s\cdot \chi) = -\omega (t\cdot \chi) \pm k (s\cdot \chi), 
\end{equation}
where the $\pm$ sign indicates the ingoing and outgoing rays, respectively. Further
\begin{equation}
k_s \equiv k_as^a=-\Omega s^T +k_r s^r; \qquad k_x \equiv k_as^a=-\Omega x^T +k_r x^r,
\end{equation} 
so that
\begin{equation}\label{krho}
k_r= \frac{k_s +\Omega s^T}{s^r} = \frac{k_x +\Omega x^T}{x^r}.
\end{equation}
Now, using the fact that $\rho$ is orthogonal to $\chi$, 
\begin{equation}
s^r= -(a\cdot \chi); \qquad x^r = -(t\cdot \chi),
\end{equation}
and we can rearrange Eq.~\eqref{krho} to see
\begin{equation}
(t\cdot \chi)= \frac{k_x- \Omega (\chi \cdot x)}{k_s- \Omega (\chi \cdot s)} (a \cdot \chi).
\end{equation}
Following closely the arguments of Appendix A of Ref. \cite{Cropp:2013sea}, for the outgoing ray,
\begin{equation}
\Omega= -\omega (t\cdot \chi) - k (x\cdot \chi) = -\omega \left(\frac{k_x- \Omega (\chi \cdot x)}{k_s- \Omega (\chi \cdot s)} (a \cdot \chi)\right) - k (s\cdot \chi).
\end{equation}
At the universal horizon $a\cdot \chi=0$ while $s\cdot \chi= |\chi|$. If everything is regular we would have $\Omega= -k|\chi|$, which cannot be true. Therefore, either $k_s= \Omega (\chi \cdot s)$, which is excluded as then $k_r=0$ or $k$, and hence $\omega$, diverge. 
The group velocity in the $t, x$ frame is simply
\begin{equation}
v_g= d\omega/dk= b_1+3bk^2,
\end{equation}
and therefore also diverges. 

Working with the four-velocity in the $t, x$ frame, $V^a= t^a +v_g x^a$, the trajectory of the ray at the universal horizon is
\begin{equation}
\frac{dT}{dr}=\frac{t^T+v_g x^T}{t^r+v_g x^r}= \frac{x^T}{x^r} \propto \frac{1}{(a\cdot \chi)}.
\label{dTdr}
\end{equation} 
and thus rays cannot escape the universal horizon; it acts as the casual barrier.

We have seen that $k$, and hence $v_g$, diverges at the universal horizon. This infinite blueshift is exactly signaling a causal barrier. However, it is worth noting that, given that the background spacetime is fundamentally Minkowski, this infinite blueshift will not be actually achieved; at very high energies our approximate dispersion relation is no longer valid, and indeed the dispersion becomes the standard relativistic one, $\omega =c k$. This breakdown of the description is intrinsic to analogue models and should not come as a surprise in a system which does provide a UV completion of the low energy effective field theory. Still it is quite intriguing the here realized scenario where in the far UV a relativity group is recovered and gravity ceases to exist.

\begin{figure}[h]
\begin{center}
\includegraphics[scale=0.8]{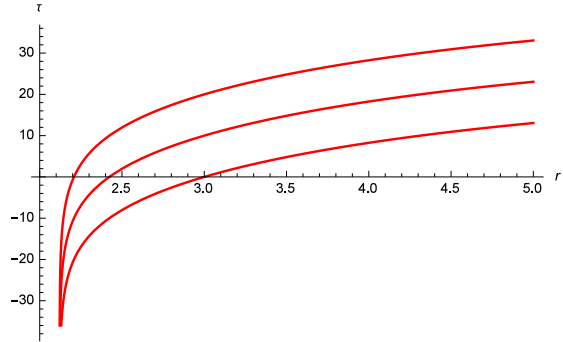}
\end{center}
\vspace{-0.5cm}
\caption{Lines of constant Khoronon, ($\Gamma$), again using the solution with $B=1$ and $m=0.1$ for Eq.~(\ref{Procafieldequations}), $R_{0}=1$ in Eq.~(\ref{unnormalizedfourvelocity}) and the speed of sound $c_{s}=0.1c$.}
\label{fig2}
\end{figure}

Further than just the blocking at the universal horizon, the \aether\ field leads to slices of constant khronon, which provides a notion of causality everywhere. In the case of time independence and spherical symmetry, given the {\aether}, we can define the khronon field by
\begin{equation}
 \Gamma=\tau+\int \frac{a_r}{a_T} dr =\tau - \int \frac{(s\cdot \chi)}{(a\cdot \chi)}, 
\end{equation}
or, in terms of the vector field $A^{\mu}$
\begin{equation}
 \tau=T+\int \frac{A^{T}g_{Tr}+A^rg_{rr}}{A^rg_{Tr}+A^{T}g_{TT}} dr.
\end{equation}
One may see that this notion of causality carries over to our analogue case also, by considering the case of $v_g \to \infty$ in Eq.~\eqref{dTdr}, corresponding to very high-energy rays. Then
\begin{equation}
\frac{dT}{dr}= \frac{x^T}{x^r} = -\frac{x_a\chi^a}{t_a\chi^a}.
\end{equation}
Up to here, we have kept things very general, but now we need to specify that the $t$ frame is defined by the velocity profile, as $v$ is precisely the natural timelike vector we have on hand, that is, $t \cdot \chi= v\cdot \chi$. Now at infinity $v=a$, and whereas $a\cdot \chi$ increases towards zero at the universal horizon, $v\cdot \chi=v_0$ decreases. This implies that 
\begin{equation}
\left|\frac{dT}{dr}\right|= \left|\frac{x\cdot \chi}{t\cdot \chi}\right| \geq \left|\frac{s\cdot \chi}{a\cdot \chi}\right|.
\end{equation}
Therefore all outgoing rays are bounded by the constant khronon surfaces, displayed in Fig.~\ref{fig2}. The fact that this is relation is an inequality, rather than an equality is precisely due to the fact that we are considering infinite-velocity rays in the dispersion frame, rather than the \aether\ frame. What is high energy in the dispersion frame can be a lower energy in the \aether\ frame and hence the ray cannot reach the limiting value of traveling along the khronon surface. 

\section{Discussion}

The purpose of this paper has been twofold. On the one side, we have considered in detail black hole solutions in the relativistic BEC framework. Given that in the hydrodynamic approximation, within which the acoustic metric is well defined, this system reproduces the equations of a relativistic perfect fluid, our results can be easily exported to this more generic setting. On the other side, we have dealt with the problem of mimicking theories with extra nonmetric structures, in particular with an \aether\ field. One of the most interesting features in these theories involves black holes and the existence of a true causal barrier, the universal horizon.  We have shown that it is possible to model such horizons by using a relativistic BEC coupled to a massive vector field. 

One key difference of relativistic BEC with respect to the standard scenario of Einstein--\Aether\ gravity is that here, even without the presence of the Proca/\aether~field, we have a modified dispersion relation. The Bose--Einstein condensation mechanism entails the breakdown of the hydrodynamical approximation at short wavelengths. The system interpolates between a low-energy limit with an emergent curved geometry for a relativistic field with a limit speed equal to the speed of sound and a high-energy limit of relativistic atoms in a flat spacetime whose causal structure is characterized by the speed of light. The interpolation between this two regimes breaks Lorentz invariance where the preferred frame is set by the condensate itself. Nonetheless, it is worth stressing that only the modification to the preferred frame introduced by the Proca field allows the presence of the universal horizon as only in this case an extra independent field other than the condensate four-velocity can exist. 

With regard to the universal horizon we have found, a comment is in order here. We have already stressed that this feature is realized only within the range of validity of the analogue gravity system at hand. Indeed, as quasiparticles are traced back towards the universal horizon, they will accumulate close to it and finally be sufficiently blueshifted so to interpolate to the dispersion relation of relativistic atoms in standard Minkowski spacetime. As such they will effectively ``disappear" from the analogue spacetime by ``melting" in the quantum substratum from which the analogue spacetime emerges. This has to be contrasted with the situation realized in the standard BEC analogue scenarios at the Killing horizon where the blueshifted traced-back quasiparticles in the end become supersonic and are able to penetrate the causal barrier. In this case the ``melting" could be a relevant issue only if the flow is set to mimic a  spacetime singularity inside the horizon.

Finally, while a rBEC system has yet to be realized in a laboratory setting, it is worth stressing that analogue models have in the past provided useful frameworks for a better understanding of theoretical issues in quantum field theory on curved spacetimes as well as emergent gravity. A typical example of this is the theoretical studies aimed at determining the robustness of Hawking radiation with respect to the nature of spacetime at ultra short scales (see e.g. Ref.~\cite{Barcelo:2005fc}). In particular, the nature of Hawking radiation in black holes with universal horizons is still debated~\cite{Berglund:2012fk, Michel:2015rsa}. From the above discussion, it seems that addressing this question would require one to address the problem of choosing suitable boundary condition (a vacuum state) at the universal horizon. In this sense it might be worth exploring numerical simulations of the system we proposed here, describing the dynamical formation of the black hole similarly to what was done in Ref.~\cite{Carusotto:2008ep} for the nonrelativistic BEC. We hope that these interesting open issues will stimulate further work on this subject.

\acknowledgments

The authors wish to thank Matt Visser for illuminating discussions and comments on the manuscript. Rodrigo Turcati is very grateful to CNPq for financial support. Bethan Cropp is supported by the Max Planck-India Partner Group on Gravity and Cosmology. This publication was made possible through the support of the John Templeton Foundation grant \#51876. The opinions expressed in this publication are those of the authors and do not necessarily reflect the views of the John Templeton Foundation.


\thebibliography{30}

\bibitem{Barcelo:2005fc} 
  C.~Barcel\'o, S.~Liberati and M.~Visser,
  ``Analogue gravity'',
  Living Rev.\ Rel.\  {\bf 8}, 12 (2005)
  [Living Rev.\ Rel.\  {\bf 14}, 3 (2011)]
  [gr-qc/0505065].

\bibitem{Garay:1999sk}
  L.~J.~Garay, J.~R.~Anglin, J.~I.~Cirac and P.~Zoller,
  ``Black holes in Bose-Einstein condensates,''
  Phys.\ Rev.\ Lett.\  {\bf 85} (2000) 4643
  [gr-qc/0002015].

\bibitem{Garay:2000jj}
  L.~J.~Garay, J.~R.~Anglin, J.~I.~Cirac and P.~Zoller,
  ``Sonic black holes in dilute Bose-Einstein condensates,''
  Phys.\ Rev.\ A {\bf 63} (2001) 023611
  [gr-qc/0005131].

\bibitem{Fagnocchi:2010sn}
  S.~Fagnocchi, S.~Finazzi, S.~Liberati, M.~Kormos and A.~Trombettoni,
  ``Relativistic Bose-Einstein Condensates: a New System for Analogue Models of Gravity,''
  New J.\ Phys.\  {\bf 12} (2010) 095012
  [arXiv:1001.1044 [gr-qc]].

\bibitem{Cropp:2013sea}
  B.~Cropp, S.~Liberati, A.~Mohd and M.~Visser,
  ``Ray tracing Einstein-Æther black holes: Universal versus Killing horizons,''
  Phys.\ Rev.\ D {\bf 89} (2014) 6,  064061
  [arXiv:1312.0405 [gr-qc]].

\bibitem{Berglund:2012fk}
  P.~Berglund, J.~Bhattacharyya and D.~Mattingly,
  ``Towards Thermodynamics of Universal Horizons in Einstein-æther Theory,''
  Phys.\ Rev.\ Lett.\  {\bf 110} (2013) 7,  071301
  [arXiv:1210.4940 [hep-th]].

\bibitem{Michel:2015rsa}
  F.~Michel and R.~Parentani,
  ``Black hole radiation in the presence of a universal horizon,''
  Phys.\ Rev.\ D {\bf 91} (2015) 12,  124049
  [arXiv:1505.00332 [gr-qc]].

\bibitem{Bilic:1999sq}
  N.~Bilic,
  ``Relativistic acoustic geometry,''
  Class.\ Quant.\ Grav.\  {\bf 16} (1999) 3953
  [gr-qc/9908002].

\bibitem{Visser:2010xv}
  M.~Visser and C.~Molina-Paris,
  ``Acoustic geometry for general relativistic barotropic irrotational fluid flow,''
  New J.\ Phys.\  {\bf 12} (2010) 095014
  [arXiv:1001.1310 [gr-qc]].



\bibitem{Berglund:2012bu}
  P.~Berglund, J.~Bhattacharyya and D.~Mattingly,
  ``Mechanics of universal horizons,''
  Phys.\ Rev.\ D {\bf 85} (2012) 124019
  [arXiv:1202.4497 [hep-th]].




\bibitem{Horava:2009uw}
  P.~Ho\v{r}ava,
  ``Quantum Gravity at a Lifshitz Point'',
  Phys.\ Rev.\ D {\bf 79} (2009) 084008
  [arXiv:0901.3775 [hep-th]].

\bibitem{Horava:2008ih} 
  P.~Ho\v{r}ava,
  ``Membranes at Quantum Criticality'',
  JHEP {\bf 0903}, 020 (2009)
  [arXiv:0812.4287 [hep-th]].

\bibitem{Visser:2009fg}
  M.~Visser,
  ``Lorentz symmetry breaking as a quantum field theory regulator'',
  Phys.\ Rev.\ D {\bf 80} (2009) 025011
  [arXiv:0902.0590 [hep-th]].




\bibitem{Blas:2011ni}
D.~Blas and S.~Sibiryakov,
  ``\Horava\ gravity versus thermodynamics: The black hole case'',
  Phys.\ Rev.\ D {\bf 84} (2011) 124043
  [arXiv:1110.2195 [hep-th]].
  
\bibitem{Barausse:2011pu}
E.~Barausse, T.~Jacobson and T.~P.~Sotiriou,
  ``Black holes in Einstein-\aether\ and \Horava--Lifshitz gravity'',
  Phys.\ Rev.\ D {\bf 83} (2011) 124043
  [arXiv:1104.2889 [gr-qc]].

 
\bibitem{Jacobson:2010mx}
 T.~Jacobson,
  ``Extended \Horava\ gravity and Einstein-\aether\ theory'',
  Phys.\ Rev.\ D {\bf 81} (2010) 101502
   [Erratum-ibid.\ D {\bf 82} (2010) 129901]
  [arXiv:1001.4823 [hep-th]].



\bibitem{Mohd:2013zca}
  A.~Mohd,
  ``On the thermodynamics of universal horizons in Einstein-{\AE}ther theory,''
  arXiv:1309.0907 [gr-qc].

\bibitem{Eling:2006ec}
  C.~Eling and T.~Jacobson,
  ``Black Holes in Einstein-Aether Theory,''
   [Class.\ Quant.\ Grav.\  {\bf 27} (2010) 049802]
  [gr-qc/0604088].

\bibitem{Sotiriou:2010wn}
  T.~P.~Sotiriou,
  ``Horava-Lifshitz gravity: a status report,''
  J.\ Phys.\ Conf.\ Ser.\  {\bf 283} (2011) 012034
  [arXiv:1010.3218 [hep-th]].

\bibitem{Blas:2009qj}
  D.~Blas, O.~Pujolas and S.~Sibiryakov,
  ``Consistent Extension of Horava Gravity,''
  Phys.\ Rev.\ Lett.\  {\bf 104} (2010) 181302
  [arXiv:0909.3525 [hep-th]].


\bibitem{Gasperini:1987nq}
  M.~Gasperini,
  ``Singularity Prevention and Broken Lorentz Symmetry,''
  Class.\ Quant.\ Grav.\  {\bf 4} (1987) 485.

\bibitem{Gasperini:1998eb}
  M.~Gasperini,
  ``Repulsive gravity in the very early universe,''
  Gen.\ Rel.\ Grav.\  {\bf 30} (1998) 1703
  [gr-qc/9805060].


\bibitem{Afshordi:2009tt}
  N.~Afshordi,
  ``Cuscuton and low energy limit of Horava-Lifshitz gravity,''
  Phys.\ Rev.\ D {\bf 80} (2009) 081502
  [arXiv:0907.5201 [hep-th]].

\bibitem{Jacobson:2013xta}
  T.~Jacobson,
  ``Undoing the twist: The Ho\v{r}ava limit of Einstein-aether theory,''
  Phys.\ Rev.\ D {\bf 89} (2014) 8,  081501
  [arXiv:1310.5115 [gr-qc]].

\bibitem{Cropp:2013zxi}
  B.~Cropp, S.~Liberati and M.~Visser,
  ``Surface gravities for non-Killing horizons,''
  Class.\ Quant.\ Grav.\  {\bf 30} (2013) 125001
  [arXiv:1302.2383 [gr-qc]].

\bibitem{Visser:1993ub}
  M.~Visser,
  ``Acoustic propagation in fluids: An Unexpected example of Lorentzian geometry,''
  gr-qc/9311028.

\bibitem{Visser:1997ux}
  M.~Visser,
  ``Acoustic black holes: Horizons, ergospheres, and Hawking radiation,''
  Class.\ Quant.\ Grav.\  {\bf 15} (1998) 1767
  [gr-qc/9712010].

\bibitem{Jacobson:2000xp}
  Jacobson T and Mattingly D 2001,
  ``Gravity with a dynamical preferred frame'',
  Phys.\ Rev.\ D {\bf 64} 024028
  [gr-qc/0007031].

\bibitem{Eling:2004dk}
  C.~Eling, T.~Jacobson and D.~Mattingly,
  ``Einstein-\AEther\ theory'',
  gr-qc/0410001.



\bibitem{Jacobson:2008aj}
  T.~Jacobson,
  ``Einstein-aether gravity: A Status report,''
  PoS QG {\bf -PH} (2007) 020
  [arXiv:0801.1547 [gr-qc]].

\bibitem{Barausse:2012ny}
  E.~Barausse and T.~P.~Sotiriou,
  ``A no-go theorem for slowly rotating black holes in Ho\v{r}ava--Lifshitz gravity'',
  Phys.\ Rev.\ Lett.\  {\bf 109} (2012) 181101
   [Erratum-ibid.\  {\bf 110} (2013) 039902]
  [arXiv:1207.6370].

    \bibitem{Barausse:2012qh}
      E.~Barausse and T.~P.~Sotiriou,
      ``Slowly rotating black holes in Ho\v{r}ava-Lifshitz gravity'',
      Phys.\ Rev.\ D {\bf 87} (2013) 087504
      [arXiv:1212.1334].


\bibitem{Kapuska:1981jj}
  J.~I.~Kapusta,
  ``Bose-Einstein condensation, spontaneous symmetry breaking, and gauge theories,''
  Phys.\ Rev.\ D {\bf 24} (1981) 426.


\bibitem{Cropp:2015tua}
  B.~Cropp, S.~Liberati and R.~Turcati,
  ``Vorticity in analogue gravity,''
  Class.\ Quant.\ Grav.\  {\bf 33} (2016) no.12,  125009
  [arXiv:1512.08198 [gr-qc]].
  
 
\bibitem{martinshaw}
  B. R.~Martin and G. Shaw,
  ``Particle Physics, 3rd edition''
  John Wiley and Sons Ltd, 2008 
  [gr-qc/9908002].

\bibitem{Liberati:2005pr} 
  S.~Liberati, M.~Visser and S.~Weinfurtner,
  ``Naturalness in emergent spacetime,''
  Phys.\ Rev.\ Lett.\  {\bf 96}, 151301 (2006)
  [gr-qc/0512139].
  
\bibitem{Jain:2007gg}
  P.~Jain, S.~Weinfurtner, M.~Visser and C.~W.~Gardiner,
  ``Analogue model of a FRW universe in Bose-Einstein condensates: Application of the classical field method,''
  Phys.\ Rev.\ A {\bf 76} (2007) 033616
  [arXiv:0705.2077 [cond-mat.other]].

\bibitem{Weinfurtner:2007dq}
  S.~Weinfurtner, A.~White and M.~Visser,
  ``Trans-Planckian physics and signature change events in Bose gas hydrodynamics,''
  Phys.\ Rev.\ D {\bf 76} (2007) 124008
  [gr-qc/0703117 [GR-QC]].
	
%


\bibitem{Carusotto:2008ep} 
  I.~Carusotto, S.~Fagnocchi, A.~Recati, R.~Balbinot and A.~Fabbri,
  ``Numerical observation of Hawking radiation from acoustic black holes in atomic BECs,''
  New J.\ Phys.\  {\bf 10}, 103001 (2008)
  [arXiv:0803.0507 [cond-mat.other]].
\end{document}